\documentclass[fleqn,10pt]{wlscirep}
\usepackage[utf8]{inputenc}
\usepackage[T1]{fontenc}
\title{Optical spin-orbit interaction induced by magnetic textures}

\author[1,2,*]{Martin Luttmann}
\author[3,1,4]{Mauro Fanciulli}
\author[5]{Pietro Carrara}
\author[5]{Maurizio Sacchi}
\author[1,+]{Thierrry Ruchon}
\affil[1]{Universit\'e Paris-Saclay, CEA, LIDYL, 91191 Gif-sur-Yvette, France}
\affil[2]{DQML, IMX, Ecole Polytechnique F\'ed\'erale de Lausanne (EPFL) Station 12, CH-1015 Lausanne, Switzerland}
\affil[3]{CY Cergy Paris Universit\'e, CEA, LIDYL, 91191 Gif-sur-Yvette, France}
\affil[4]{New Technologies Research Center, University of West Bohemia, 30100 Plze\v{n}, Czech Republic}
\affil[5]{Sorbonne Université, CNRS, Institut des NanoSciences de Paris, INSP, F-75005 Paris, France}

\affil[*]{martin.luttmann@epfl.ch}
\affil[+]{thierry.ruchon@cea.fr}

\begin{abstract}
Contrary to the optical spin angular momentum (SAM), the role played by the orbital angular momentum (OAM) of light in magneto-optics remains largely unexplored. However, the SAM and OAM are known to be coupled when light interacts with non-homogeneous and non-isotropic materials. Here we predict that the OAM carried by each photon in a light beam is modified upon reflection on magnetic textures like skyrmions, and that the sign of this variation is governed by the SAM of the incident field. Our predictions can be readily tested by performing circular dichroism measurements, and may provide new routes to shape light's angular momentum with magnetism.
\end{abstract}
\begin{document}

\flushbottom
\maketitle
% * <john.hammersley@gmail.com> 2015-02-09T12:07:31.197Z:
%
%  Click the title above to edit the author information and abstract
%
\thispagestyle{empty}

\section*{Introduction}

Spin-orbit interaction (SOI) is a phenomenon occurring in many areas of quantum physics, for instance in condensed matter in the case of noncollinear magnetism \cite{Sandratskii1998}, magnetocrystalline anisotropy \cite{Daalderop1990}, Rashba \cite{Bychkov1984} and Dresselhaus effects \cite{Dresselhaus1955}, anomalous Hall effect \cite{Nagaosa2010}, spin Hall effects \cite{Sinova2015}, topological materials \cite{Hasan2010} and the electronic structure of heavy element compounds such as transition metal dichalcogenides \cite{Zhu2011}, or in dilute matter for gas phase spectroscopy \cite{Zhong2020}.
More recently, this concept was extended to the spin and orbital angular momenta (SAM and OAM) of light, opening the field of \textit{optical spin-orbit interactions} (OSOI). Through OSOI, light angular momentum of one type -- spin or orbital -- may be controlled by the value of the other type \cite{Bliokh2015NatPhot}; in this respect, OSOI is a quite distinct concept from electronic SOI in condensed-matter systems.
OSOI has been identified upon strong focusing conditions, or through refraction or diffraction. However, OSOI was also reported when light propagates in a medium that is both anisotropic and inhomogeneous, with important applied perspectives for the control of the SAM and OAM of light. This is the working principle of optical elements like q-plates \cite{Marrucci2006,Karimi2009} (exhibiting spatially varying birefringence), metasurfaces \cite{Yin2013, Devlin2017} or light-shaping nanostructures \cite{Bomzon2002,Brasselet2009}. These examples all rely on a structural patterning of matter, possibly controlled by electric fields. On the contrary, while the coupling between light polarization and magnetism is well established \cite{Schatz1969, Ebert1996, Arregi2016, Stephens1974, Funk2005}, the research effort aiming at identifying the role of OAM in magneto-optical interactions is only very recent \cite{Fujita2017, Yang2018, Sirenko2019, Fanciulli2022, Fanciulli2025} and studies of OSOI induced by magnetism are very scarce today \cite{Levy2017, Lei2021}. However, as the ensemble of stable magnetic structures, with fine control on their manufacturing and innumerable applications, is growing, it is timely to investigate their specific impact on light beams carrying SAM and OAM. Some promising textures like magnetic  vortices and skyrmions \cite{Skyrme1961, Bogdanov1989, Muhlbauer2009,Soumyanarayanan2017,Gobel2021} are particularly attractive for these investigations. They are patterned in dots of swirling magnetization, while their typical size ranges from a few microns, to a few nanometers, which is comparable to focal spot sizes of visible to soft X-ray beams, a spectral range where light beams with angular momentum are now available \cite{Geneaux2016, Kong2017}. In this context, identifying magnetically-induced OSOIs might open interesting perspectives, for instance for the ultrafast and microscopic control of the angular momentum of light. 

Here we harness the theoretical framework proposed in Ref. \cite{Fanciulli2021} to demonstrate how the average OAM per photon in a light beam can be modified upon reflection on magnetic textures. 
In particular, we show that the sign of the OAM variation is controlled by the SAM of the incident field, while its magnitude depends on the specific magnetic structure considered. This magnetically-induced OSOI originates from the azimuthal interference between the longitudinal, transverse and polar magneto-optical Kerr effect (MOKE) components, that get coupled for incident elliptically polarized light fields. Taking the example of the X-ray spectral range, we show that magnetic vortices and skyrmions are an ideal test-bed for the observation of this form of OSOI. 

\section*{Results}

\subsection*{Magnetic OSOI}
\label{sec: MO model}

The notations introduced in Ref.\,\cite{Fanciulli2021} will be used throughout this paper, with simplifications when possible; we will note $\ell$ the topological charge of the beam (corresponding to the average OAM), and $s$ the SAM of the beam. For a pure mode of OAM $\ell$ and SAM $s$, they correspond to the OAM and SAM of individual photons divided by $\hbar$.
We consider  an incident vortex beam \cite{Allen1992, Coullet1989, Padgett2017} that is a pure mode of OAM quantum number $\ell_{\text{in}}$. It writes, in the $(p, s)$ polarization basis, as 
\begin{equation}
	\vec{E}_{\text{in}}=A(r,z)e^{i\ell_{\text{in}}\phi}\begin{pmatrix}
		\epsilon_p\\
		\epsilon_s
	\end{pmatrix}\,.
	\label{eq:LGBeam}
\end{equation}
$A(r,z)$ is a scalar function, with radial ($r$) and longitudinal ($z$) dependence, which can for instance correspond to a Laguerre-Gaussian (LG) amplitude \cite{Allen1992}. $\phi$ is the azimuthal angle, as measured with respect to the center of the beam. $(\epsilon_p, \epsilon_s)$ is a unit vector describing the polarization state of the incident field, $(1, 0)$ and $(0, 1)$ corresponding to $p$-polarization and $s$-polarization, respectively. Circular polarization is expressed as $(\epsilon_p, \epsilon_s) = \frac{1}{\sqrt{2}}(1, -i\,s_{\text{in}})$, $s_{\text{in}} = +1$ and $-1$ corresponding respectively to circular right (CR) and circular left (CL) polarization.
The incident beam is reflected by a planar magnetic sample with three-dimensional magnetization. Considering the linear MOKE, the reflected field $\vec{E}_{\text{out}}$ is given by the product of $\vec{E}_{\text{in}}$ with the  reflectivity matrix \cite{Arregi2016, Qiu1999}
\begin{equation}
	\mathbf{R}=\begin{pmatrix}
		r_{pp}\cdot\left[1+ r_0^t  \cdot m^t\right]& r_{ps}^l \cdot m^l+ r_{ps}^p\cdot m^p \\
		- r_{ps}^l\cdot m^l+ r_{ps}^p\cdot m^p & r_{ss}
	\end{pmatrix}\,,
	\label{eq:reflectionMatrixA}
\end{equation}
where $m^t$, $m^l$, $m^p$ are the transverse, longitudinal and polar components of the magnetization with respect to the plane of incidence, normalized by the saturation magnetization. $r_{pp}$ and $r_{ss}$ are the standard Fresnel reflectivity coefficients and $r_0^t $, $r_{ps}^l$ and $r_{ps}^p$ are magneto-optical constants. They depend on the material, the wavelength and the incidence angle \cite{Arregi2016, Piovera2013}. As we consider an inhomogeneous sample, it should be noted that this reflectivity matrix generally depends on both $r$ and $\phi$.

To simplify the analytical derivations, as in Ref. \cite{Fanciulli2021}, we assume a close to normal incidence, so that we can identify the coordinates in the transverse plane of the beam and in the sample plane. Moreover, we drop the radial dependence of the magnetization of simplicity, as it is not expected to have an impact on the OAM content. Note that we will use in our analytical calculations the magneto-optical constants computed at Brewster's angle. Comparing to realistic simulation results far from normal incidence in the following section, we will confirm the soundness of these approximations to capture the essence of magnetically-induced OSOI. Since the magnetization must be $2\pi$ periodic with $\phi$, its components can be expressed as azimuthal Fourier series \cite{Fanciulli2021}
\begin{equation}
\label{eq:azimuthal Fourier series}
    m^\dag(\phi) = m_0\sum_{n=-\infty}^{+\infty}a_n^\dag e^{in\phi}\,,
\end{equation}
where the $\dag = t,l,p$ can refer to each of the three components of the magnetization, and where $m_0$ is a dimensionless quantity. %Without loss of generality, we can consider that a non-null $a_0^\dag$, which would correspond to a homogeneous magnetization part, could be included in the regular Fresnel coefficients $r_{pp}$ and $r_{ss}$ and we then set $a_0^\dag=0$.  
The field after reflection off the magnetized material writes, using Eq. 	\ref{eq:LGBeam}, 	\ref{eq:reflectionMatrixA} and \ref{eq:azimuthal Fourier series},
%\begin{widetext}
\begin{align}
\vec{E}_{\text{out}} &= \textbf{R} \cdot \vec{E}_{\text{in}}\nonumber\\
& =A(r,z)e^{i\ell_{\text{in}}\phi}
\begin{pmatrix}
\epsilon_pr_{pp}\cdot\left[ 1+ r_0^t  m^t\right] + \epsilon_s\cdot\left[r_{ps}^l m^l + r_{ps}^p m^p \right] \\
\epsilon_p\cdot\left[ -r_{ps}^l m^l + r_{ps}^p m^p \right] +\epsilon_s r_{ss}
\end{pmatrix}\nonumber\\
&=A(r,z)		\begin{pmatrix}
			\epsilon_p r_{pp} e^{i\ell_{\text{in}}\phi} + m_0 \sum_n\left[\epsilon_p r_{pp}r_0^t a_n^t + \epsilon_s\left( r_{ps}^l a_n^l + r_{ps}^p a_n^p \right) \right] e^{i(\ell_{\text{in}}+n) \phi} \\
			\epsilon_s r_{ss} e^{i\ell_{\text{in}}\phi} + m_0 \sum_n \epsilon_p \left( -r_{ps}^l a_n^l + r_{ps}^p a_n^p \right) e^{i(\ell_{\text{in}}+n) \phi}
		\end{pmatrix}\,.
        \label{eq:reflection of an gaussian CR beam}
\end{align}
For a general incident elliptical polarization state, terms corresponding to transverse, longitudinal and polar MOKE contribute to the $p$-polarized component of the reflected field, while the $s$-polarized component has contributions from longitudinal and polar MOKE only.
We also notice that the OAM $\ell_{\text{in}}$ of the incident field has been redistributed over the OAM modes $\ell_{\text{in}}+n$, the weight of each mode being proportional to $\left|\epsilon_p r_{pp}r_0^t a_n^t + \epsilon_s\left( r_{ps}^l a_n^l + r_{ps}^p a_n^p \right)\right|^2 + \left| \epsilon_p \left(-r_{ps}^l a_n^l + r_{ps}^p a_n^p \right)\right|^2$, for $n \neq 0$. From Eq.\,\ref{eq:reflection of an gaussian CR beam} we see that a uniform magnetization (i.e. the $a_0^{\dag}$ terms) contributes, as expected, to mixing the $p$ and $s$ components, while it does not contribute to OAM redistribution nor to OSOI. As we are interested in the variation of OAM, we set $a_0^\dag = 0$ in what follows, namely we only consider inhomogeneous magnetization patterns.
We denote respectively $\mathcal{W}_{\text{in}}$ and $\mathcal{W}_{\text{out}}$ the energies in the incoming and outgoing beams. It can be shown that (see Eq.\,S42 of the supplementary material)
\begin{equation}
    \mathcal{W}_{\text{out}} =\mathcal{W}_{\text{in}}\cdot\left[ \left|\epsilon_p r_{pp}\right|^2 + \left|\epsilon_s r_{ss}\right|^2 + |m_0|^2 \sum_{n \neq 0} \left(  \left|\epsilon_p r_{pp}r_0^t a_n^t +\epsilon_s \left(r_{ps}^l a_n^l + r_{ps}^p a_n^p \right) \right|^2 + \left|\epsilon_p \left( -r_{ps}^l a_n^l + r_{ps}^p a_n^p \right)\right|^2 \right)\right]\,.
    \label{eq:intensity}
\end{equation}
The mean OAM quantum number for the reflected field is (Eq.\,S43 of the supplementary material) 
\begin{equation}
\label{eq: mean OAM variation}
    \ell_{\text{out}} = \ell_{\text{in}} + |m_0|^2\cdot \frac{\mathcal{W}_{\text{in}}}{\mathcal{W}_{\text{out}}} \cdot 
     \sum_{n \neq 0} n \left( \left| \epsilon_p r_{pp}r_0^t  a_n^t + \epsilon_s \left( r_{ps}^l a_n^l + r_{ps}^p a_n^p \right)\right|^2 + \left|\epsilon_p(-r_{ps}^l a_n^l + r_{ps}^p a_n^p)\right|^2  \right)\,.
\end{equation}
Similarly, the mean SAM quantum number in the outgoing beam, projected on the propagation axis, reads (Eq.\,S45)
\begin{equation}
\label{eq: mean SAM variation}
	\begin{split}
		s_{\text{out}}	&=2\frac{\mathcal{W}_{\text{in}}}{\mathcal{W}_{\text{out}}}\cdot\left[\mathcal{I}\left(\epsilon_p^*\epsilon_s r_{pp}^* r_{ss} \right)+
			|m_0|^2\cdot\sum_{n\neq 0}{\mathcal{I}\bigg(\left[ \epsilon_p r_{pp} r_0^t  a_n^t + \epsilon_s \left(r_{ps}^l a_n^l+r_{ps}^p a_n^p\right)\right]^*\cdot \epsilon_p \left( -r_{ps}^l a_n^l+r_{ps}^p a_n^p\right) \bigg)}	\right]	\,.
		\\
	\end{split}
\end{equation}
%\\end{widetext}

At this point, several conclusions can been drawn. First, in the case of a homogeneous magnetization, which corresponds to $a_n^\dag=0$ for $n\neq 0$, we see that there is no change of OAM, as anticipated from Eq.\ref{eq:reflection of an gaussian CR beam}. % and that the SAM is only affected by the Fresnel reflectivity coefficients (first term on the right hand side of Eq. \ref{eq: mean SAM variation}). 
Second, for an inhomogeneous magnetization we find that neither the OAM nor the SAM changes depend on the magnetization direction, i.e. on the sign of $m_0$, but rather on its modulus square and on the relative weights of its Fourier components. %Third, a non-null incoming SAM, which corresponds to $\mathcal{I}(\epsilon_p\epsilon_s^*)\neq 0$ may cause variation of the outgoing OAM (first quadratic term in the right handside of Eq. \ref{eq: mean OAM variation}), but the OAM of the incoming field does not affect the SAM of the outgoing field. 
Third, the OAM of the incoming field does not affect the SAM of the outgoing field (Eq.\,\ref{eq: mean SAM variation}), while the opposite may not be true (Eq.\,\ref{eq: mean OAM variation}). Indeed, as we will show later, a non-null incident SAM will cause an OAM variation. We identify this OAM change, controlled by the SAM, to a magneto-optical SOI. Finally, we conclude from Eq.\,\ref{eq: mean OAM variation},\ref{eq: mean SAM variation} that in general the variation of OAM is not compensated by an opposite variation of SAM. This is expected, as a non-zero incidence angle as well as a spatially varying magnetization may both break the rotational symmetry of the problem.

The formulae given in Eq.\,\ref{eq:intensity}, \ref{eq: mean OAM variation} and \ref{eq: mean SAM variation} greatly simplify in cases of high rotational symmetry, such as magnetic vortices or skyrmions. In the following calculations, we will consider a simple planar magnetic vortex \cite{Fanciulli2021}, which approximates a Bloch skyrmion (a realistic skyrmion texture will be used for the simulations in the following section). In this case, there is no azimuthal variation of $m^p$, and the azimuthal variations of $m^t$ and $m^l$ are purely harmonic, with periodicity $2\pi$; we thus have $a_n^p=0$, and only the $n=\pm 1$ terms contribute to  $a_n^t$ and $a_n^l$: 
\begin{equation}
	\label{eq:vortex}
	\begin{split}
		a_n^t &= 
		\begin{cases}
			\frac{1}{2}, \, n = \pm 1 \\
			0, \, n \neq \pm 1
		\end{cases}\\
		a_n^l &= 
		\begin{cases}
			n\frac{i}{2}, \, n = \pm 1 \\
			0, \, n \neq \pm 1
		\end{cases}\\
		a_n^p &= 0
	\end{split}
    \,.
\end{equation}
 %It should be noted that with this degree of approximation, the Bloch skyrmions is very much similar to a magnetic vortex \cite{Fanciulli2021}. 
 The reflected beam is in a superposition of modes $\ell_{\text{in}}$, $\ell_{\text{in}}+1$ and $\ell_{\text{in}}-1$ and the change in OAM reduces to
\begin{align}
  \ell_{\text{out}} - \ell_{\text{in}} &=|m_0|^2\frac{\mathcal{W}_{\text{in}}}{4\mathcal{W}_{\text{out}}}\cdot \bigg( \left|\epsilon_p r_{pp}r_0^t  + i\epsilon_sr_{ps}^l\right|^2 - \big.\nonumber
\big.\left|\epsilon_p r_{pp}r_0^t  - i\epsilon_sr_{ps}^l\right|^2  \bigg)  \nonumber\\
  &=|m_0|^2\frac{\mathcal{W}_{\text{in}}}{\mathcal{W}_{\text{out}}}\cdot\mathcal{I} \bigg( \epsilon_p \epsilon_s^* \cdot r_{pp}r_0^t \cdot r_{ps}^{l,*} \bigg)\,.
  \end{align}
  
\begin{figure}
	\centering
	%\resizebox{0.8\the sign o the textwidth}{!}
	\includegraphics[width=0.25\textwidth]{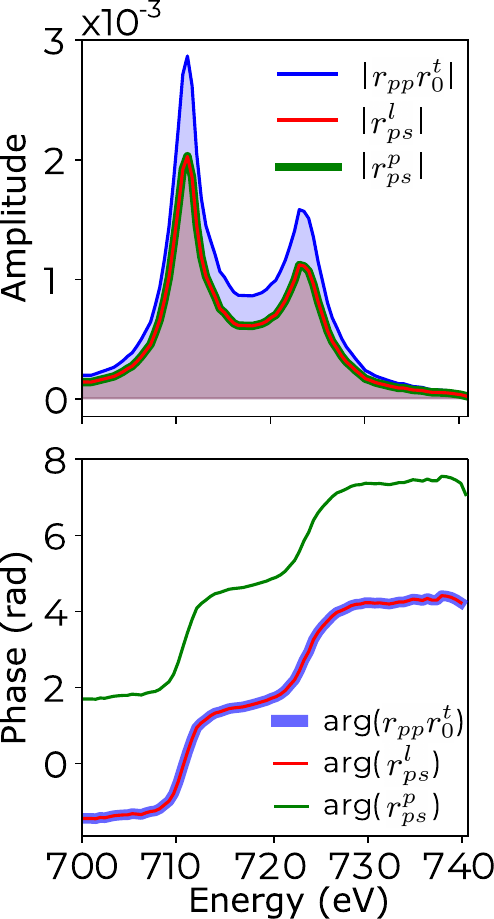}
	%}
\caption{Amplitude (top) and phase (bottom) of the magneto-optical constants $r_{pp}r_0^t $, $r_{ps}^l$ and $
	r_{ps}^p$ at the L-edge of iron, for an angle of incidence of 45\textdegree.}
\label{fig:Moconstants}
\end{figure}

In practice, $r_{pp}r_0^t $ and $r_{ps}^l$ constants may often have the same phase. As a practical example, we consider the L$_3$-edge of Fe, corresponding to a photon energy of 711 eV, and a wavelength of about 1.7 nm (see Section I of the supplementary material for the derivation of the constants). Fig.\,\ref{fig:Moconstants} shows the different magneto-optical constants at Brewster's angle, with respect to the photon energy. We observe that the three constants have similar amplitudes. We also note that $r_{pp}r_0^t $ and $r_{ps}^l$ have the exact same complex phase, but are $\pi$ out of phase with $r_{ps}^p$. Taking theses phases into account, in the region of the L$_3$-edge of Fe we finally get 
\begin{equation}
		\ell_{\text{out}} - \ell_{\text{in}} =|m_0|^2 \frac{\mathcal{W}_{\text{in}}}{\mathcal{W}_{\text{out}}}\cdot\mathcal{I} \big(\epsilon_p\epsilon_s^* \big)\cdot \left|r_{pp}r_0^t r_{ps}^{l,*}\right|\,.  
\end{equation}
At this point, we can differentiate two cases, according to the incident polarization.

\begin{figure*}[hbt!]
	\centering
	%\resizebox{0.8\the sign o the textwidth}{!}
	\includegraphics[width=0.8\textwidth]{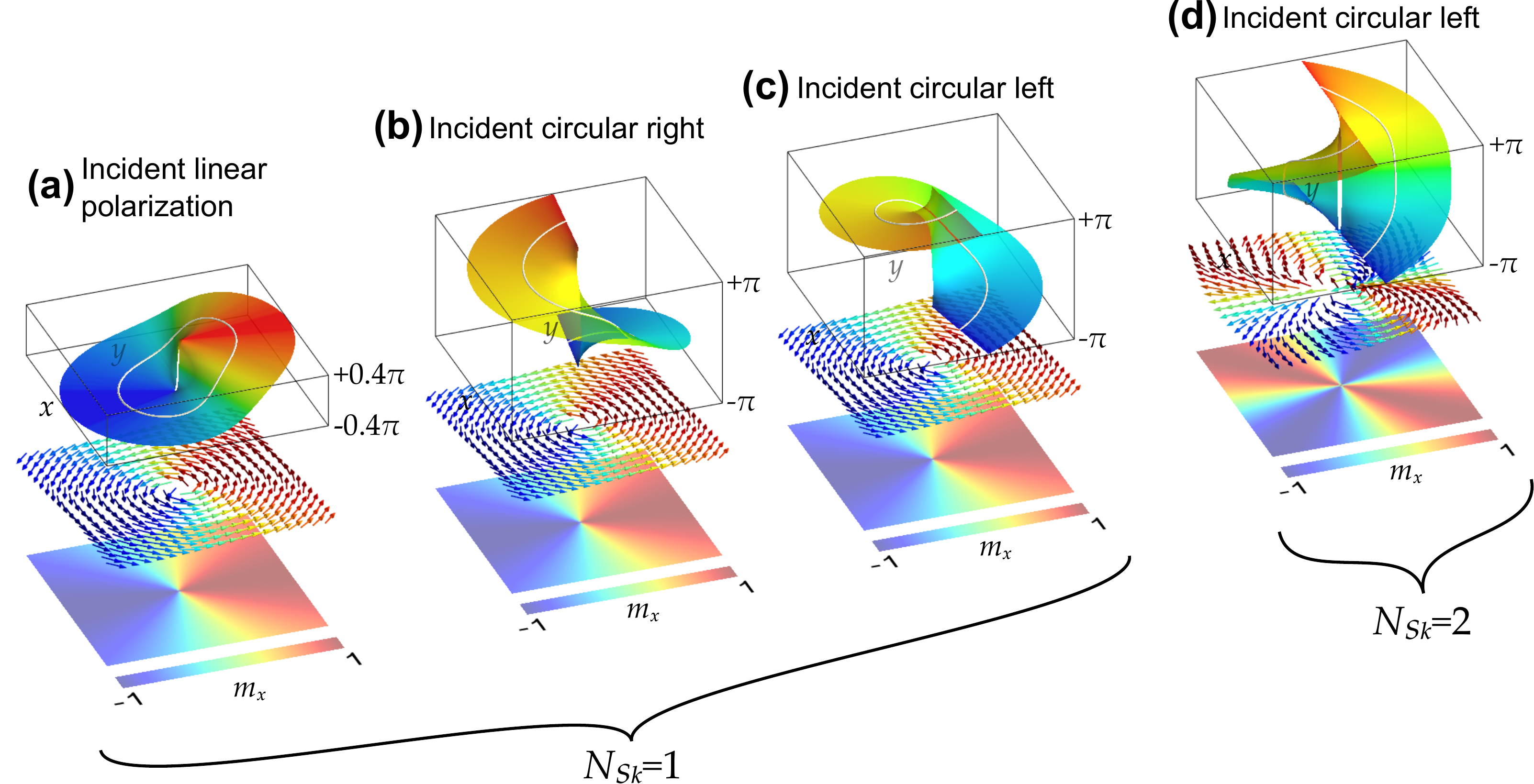}
	%}
\caption{Spatial dephasing of the $p$ component of the field upon reflection off a vortex of swirling magnetization. The isosurface color indicates the local phase of the reflected beam, while the bottom images show the transverse magnetization magnitude $m_x$. (a) For an incident $p$-polarized field, the variation of OAM is null, as the net dephasing along a loop about the optical axis (white line) is null. (b,c) For an incident CR ($s_{\text{in}}=1$) and CL ($s_{\text{in}}=-1$) field. There is a $2\pi$ dephasing along the loop, thus a positive (b) or negative (c) OAM variation. Higher-order vortices induce higher variations of the OAM (d). Here we obtain a $4\pi$ azimuthal phase variation with a magnetic vortex of order $N_{\text{Sk}}=2$.}
\label{fig:SOI}
\end{figure*}

If the incident light is linearly polarized, which corresponds to $\epsilon_p$ and $\epsilon_s$ having the same phase, one finds $\ell_{\text{out}} = \ell_{\text{in}}$.  Fig.\,\ref{fig:SOI}.a shows the kick of transverse phase of the $p$ component of the light upon reflection [i.e. the phase of $\epsilon_p(r_{pp} + r_{pp}r_0^t m^t) + \epsilon_s(r_{ps}^l m^l + r_{ps}^p m^p)$], for a planar magnetic vortex and a $p$-polarized incident field. While the magneto-optical interaction does induce a curvature of the beam wavefront, the mean OAM is unchanged, as the accumulated phase along a loop surrounding the optical axis is zero.

If the incident light is circularly polarized, we find 
\begin{equation}
    \ell_{\text{out}} - \ell_{\text{in}} =  s_{\text{in}} \frac{\mathcal{W}_{\text{in}}}{\mathcal{W}_{\text{out}}}\cdot\frac{ |m_0|^2}{2}\left|r_{pp}r_0^t r_{ps}^l\right| \,.
    \label{eq:mean OAM variation vortex CP}
\end{equation}
We thus unveil a form of OSOI, since the OAM change can be positive or negative depending on the helicity of the incident field $s_{\text{in}}$ (Fig.\,\ref{fig:SOI}.b,c). Returning to Eq.\,\ref{eq:reflection of an gaussian CR beam}, we note that this change of OAM only occurs in the $p$-polarized component of the beam, whereas the $s$-polarized component retains an OAM equal to $\ell_{\text{in}}$. Indeed, the $p$ component of the reflected field contains the interference between the transverse and longitudinal magnetization terms, coupled through the presence of both $\epsilon_p$ and $\epsilon_s$, resulting in a variation of OAM. On the other hand, the $s$ component of the reflected field is simply given by a term proportional to the longitudinal magnetization, up to a constant term. Note however that in the presence of azimuthally-varying longitudinal \textit{and} polar magnetization components, an OAM variation can also occur in the $s$ component of the light field. 
We can also compute the mean outgoing SAM quantum number as 
\begin{align}
	s_{\text{out}}&=2
\frac{\mathcal{W}_{\text{in}}}{\mathcal{W}_{\text{out}}}\left[	\mathcal{I}\left(\epsilon_p^*\epsilon_s r_{pp}^*r_{ss} \right)- \frac{|m_0|^2}{2} \mathcal{I}\left(\epsilon_p \epsilon_s^*\right) \left|r_{ps}^l\right|^2	\right]\nonumber\\
	&=\frac{\mathcal{W}_{\text{in}}}{\mathcal{W}_{\text{out}}}\left[-s_{\text{in}}	\mathcal{R}\left( r_{pp}^*r_{ss} \right) + s_{\text{in}} \frac{|m_0|^2}{2} 	\left|r_{ps}^l\right|^2  \right]\,.
	\label{eq:SAMMultimodeAveraged1}
\end{align}
The first term on the right hand side is the well-known change of polarization upon reflection on a surface due to the dephasing between the $s$ and $p$ polarization components. For instance, setting $r_{pp}=-r_{ss}=1$, which would model a perfect metallic mirror at 45$^{\circ}$ incidence, we retrieve the expected inversion of SAM. We notice the strong analogy between the second term and the OAM change in Eq. \ref{eq:mean OAM variation vortex CP}, which are similar up to the replacement of one  $\left|r_{ps}^l\right|$ term by $\left|r_{pp}r_0^t\right|$. However, the difference in magnitude prevents them from compensating one another, and we conclude that the total angular momentum of light is not conserved during the interaction. This results from the fact that the sample is not at a normal angle with respect to the incident beam's propagation axis, and thus does not show rotation symmetry about this axis. This is again analog to non-rotationally symmetric q-plates, which do not allow a pure SAM to OAM conversion \cite{Marrucci2006}.

\subsection*{Simulation of the magneto-optical interaction}

\label{sec: simu}

To test our predictions, we perform simulations of a realistic MOKE experiment, not at normal incidence, using the magneto-optical constants at the L$_3$-edge of Fe. The photon energy of the incident beam is 711.2 eV, and the incidence angle is 45$^\circ$ (Brewster's angle). We express the spin texture of the magnetic skyrmions using the following ansatz \cite{Brearton2020, Leon2016}:
\begin{equation}
	\label{eq: skyrmion}
	\begin{split}
		\textbf{m}_{\text{Sk}} &= (m^l, m^t, m^p)\\
		&= \bigg( C(\phi) \sin\{f(r)\}, S(\phi) \sin\{f(r)\},  \cos\{f(r)\} \bigg) \,,
	\end{split}
\end{equation}
with $C(\phi)=\cos(N_{\text{Sk}}\phi+h)$ and $S(\phi)=\sin(N_{\text{Sk}}\phi+h)$. The helicity $h$ determines the type of the skyrmion. $h = (0, \pi)$ corresponds to Néel-type solutions (with zero azimuthal magnetization component) and
$h =\pm \pi/2$ yields Bloch-type solutions (a magnetization vortex). $N_{\text{Sk}}$ is the skyrmion charge, equal to 1 for the Bloch and Néel skyrmions considered here. We will later comment on the case of textures with $N_{\text{Sk}}>1$. A good approximation of the radial function $f(r)$ is \cite{Leon2016}
\begin{equation}
	f(r) =4 \tan^{-1}\big(e^{-\frac{r}{\lambda}} \big) \,,
\end{equation}
where $\lambda$ is the extension of the skyrmion. In numerical applications, we will set $\lambda = 100$ nm, in order to match the size of typical skyrmions. The incident beam waist is also set to 100 nm. The other parameters of the simulation (in particular the exact modelling of magnetic sample and substrate) are the same as in Ref.\,\cite{Fanciulli2021,Fanciulli2022}. The reflected field at the sample's plane (Eq. \ref{eq:reflection of an gaussian CR beam}) is numerically propagated to the far field using the Fraunhofer operator. 

%The time-averaged energy density of a monochromatic electromagnetic field of frequency $\omega$ is given by
%\begin{equation}
%    W = \frac{g \omega}{2} (|\textbf{E}|^2+|\textbf{H}|^2),
%\end{equation}
%where $g=\frac{1}{8\pi \omega}$ in gaussian units, and $\textbf{E}$ and $\textbf{H}$ are the electric and magnetic fields. The canonical momentum density can be expressed as
%\begin{equation}
%    \textbf{P} = \frac{g}{2} \mathcal{I} \big[\textbf{E}^*\cdot(\nabla)\textbf{E} + \textbf{H}^*\cdot(\nabla)\textbf{H} \big],
%\end{equation}
%where $\mathcal{I}$ stands for the imaginary part, and where 
%\begin{equation}
%\textbf{E}^*\cdot(\nabla)\textbf{E} = 
%\begin{pmatrix}
%\sum_j \epsilon_j^* \frac{\partial \epsilon_j}{\partial_x} \\
%\sum_j \epsilon_j^* \frac{\partial \epsilon_j}{\partial_y}\\
%\sum_j \epsilon_j^* \frac{\partial \epsilon_j}{\partial_z}
%\end{pmatrix}.
%\end{equation}
%Finally, the density of OAM and SAM projected along the $z$ axis are respectively given by 
%\begin{equation}
%    L_z = (\textbf{r} \times \textbf{P}) \cdot \textbf{u}_z,
%\end{equation}
%\begin{equation}
%    S_z = \frac{g}{2} \mathcal{I} \big[\textbf{E}^*\times \textbf{E} + \textbf{H}^*\times\textbf{H} \big] \cdot %\textbf{u}_z,
%\end{equation}
%where $\textbf{u}_z$ is a unit vector along the $z$ axis.

\begin{figure}
\centering
%\resizebox{0.8\the sign o the textwidth}{!}
\includegraphics[width=0.48\textwidth]{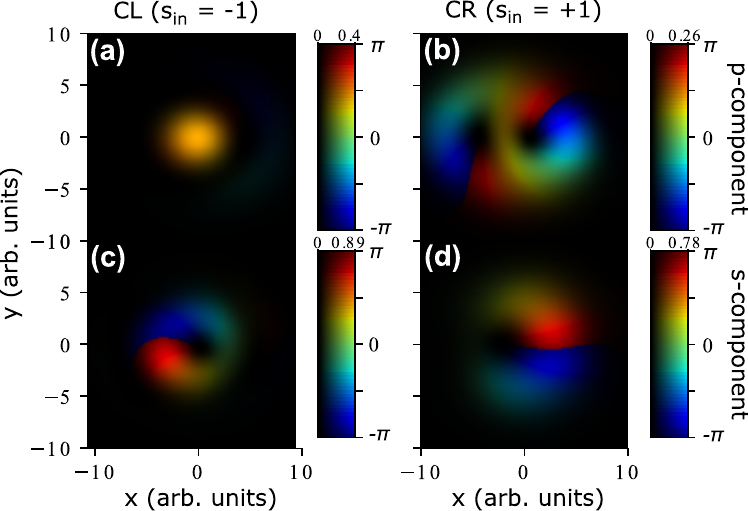}
%}
\caption{Analysis of the reflected beam in the far field, for a CL (left column) and CR (right column) incident field with $\ell_{\text{in}}=1$, impinging on a Bloch skyrmion. (a,b) $p$-polarization component. (c,d) $s$-polarization component. The phase is indicated by the color, and the local intensity corresponds to the brightness of the image. }% (e,f) OAM density variation upon reflection. (g,h) SAM density. The far field is normalized so that the peak of the total intensity is 1.}
 \label{fig:simu SOI}
\end{figure}

Fig.\,\ref{fig:simu SOI} displays the results for an incident LG mode of OAM $\ell_{\text{in}}=1$, with $s_{\text{in}}=\pm1$, impinging on a Bloch skyrmion. For $s_{\text{in}}=-1$, inspecting the intensity and phase of the $p$ and $s$ components of the reflected field at the detector (Fig.\,\ref{fig:simu SOI}.a,c), we observe that the $p$ component has been almost fully converted into the $\ell=0$ mode, as it exhibits a strong on-axis intensity and flat spatial phase. As expected, the $s$ component retains the $\ell=1$ helical phase of the incident beam. On the other hand, for $s_{\text{in}}=+1$ an increase in OAM is manifest in the phase of the $p$ component, which shows a $4\pi$ azimuthal variation, typical of an OAM of 2 (Fig.\,\ref{fig:SOI}.b). Similar observations can be made for a Néel-type skyrmion (Fig.\,S1 of the supplementary material). The reflected field is a superposition of different modes, and thus does not exhibit the perflectly round, donut-like profile typical of pure LG modes.

\begin{figure}[htbp]
\centering
%\resizebox{0.8\the sign o the textwidth}{!}
\includegraphics[width=0.47\textwidth]{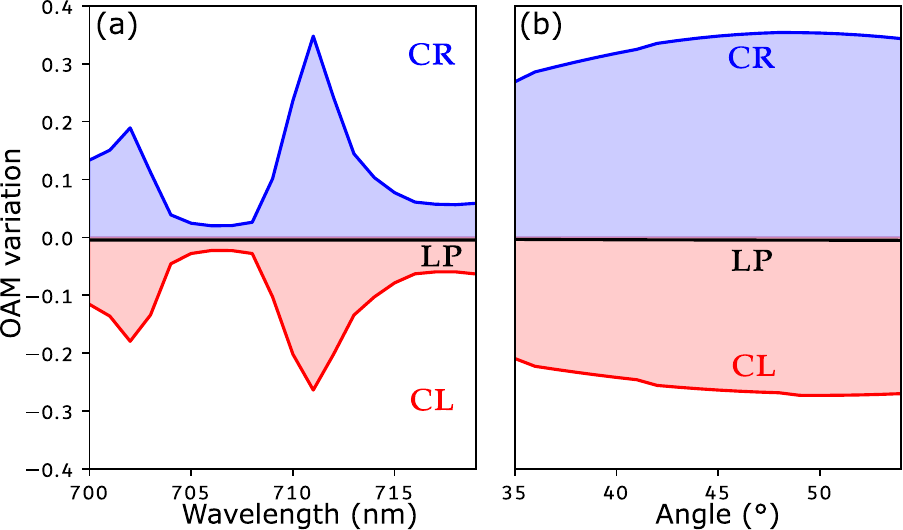}
%}
\caption{OAM variation per photon $\Delta \ell$  upon reflection on a Bloch skyrmion, with respect to the photon energy at Brewster's angle (a), and with respect to the incidence angle (in degree from normal incidence) at the L-edge of Fe, for incident CR (blue), CL (red) and LP (black) polarization.}% The dotted lines indicate, with arbitrary units, the corresponding total reflected intensity for each case.}
 \label{fig:map OAM var}
\end{figure}

To get a more quantitative insight of the OAM variation, we compute numerically the local OAM value in the far field using Eq.\,S24 (see Fig.S1.e-h). This quantity is then weighted by the local intensity and averaged over the beam profile, yielding the average OAM value per photon in the reflected beam. Subtracting the incident OAM $\ell_{\text{in}}$, we obtain the mean OAM variation $\Delta \ell = \ell_{\text{out}} - \ell_{\text{in}}$ caused by the reflection on the skyrmion. $\Delta \ell$ is shown as a function of the photon energy and the incidence angle in Fig.\,\ref{fig:map OAM var}. As expected, we find that in the vicinity of the L-edge of Fe, around 710 eV, $\Delta \ell$ is positive for a CR incident polarization ($s_{\text{in}} =+1$) and negative for a CL incident polarization ($s_{\text{out}} =-1$). In particular, at the resonance frequency and at the Brewster's angle, we find $\Delta \ell \approx \pm 0.3$. We attribute the slight asymmetry between the CR and CL curves to numerical errors, although we have not entirely identified its origin so far.
%, to be compared to the $\pm 40$ \% predicted by our analytical derivation.  %These observations are in full agreement with the conclusions of our analytical model stating that the SAM of the incident field controls the sign of the topological charge change. 
We conclude that, as anticipated with strong approximations in the analytical section, the SAM of the incident field indeed controls the sign of the OAM change. 

So far we considered textures with charge $N_{\text{Sk}}=1$. However our calculations and simulations are also valid for higher values of the skyrmion charge. Higher-order skyrmions, with arbitrary charge $N_{\text{Sk}}$ \cite{Gobel2021}, will induce a variation of the OAM of the $p$ component of the field equal to $N_{\text{Sk}}$ in circularly polarized light (Fig.\,\ref{fig:SOI}.d). 
Careful engineering of the magnetic topology thus allows fine control of the OAM of the reflected beam. %We conclude that the absorption of the incident light by the material, and the change of light's SAM upon reflection, are not the only sources of light-matter angular momentum exchange here. The modification of the average orbital angular momentum per photon also participates in a transfer of positive of negative angular momentum from the light beam to the magnetic sample.

\subsection*{On-axis magnetic circular dichroism}

\begin{figure}
\centering
%\resizebox{0.8\textwidth}{!}
\includegraphics[width=0.45\textwidth]{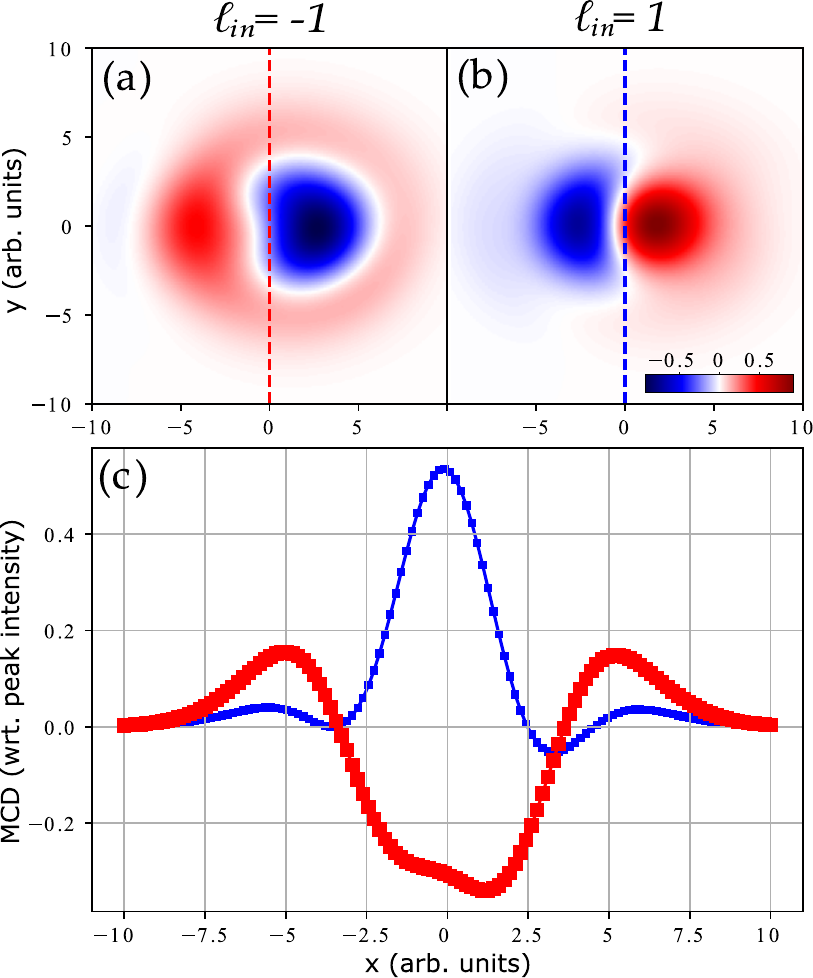}
%}
\caption{Magnetic circular dichroism images for a $h= \pi/2$ Bloch skyrmions (top images), and for an incident OAM $\ell_{\text{in}}=-1$ (a, b) and $\ell_{\text{in}}=1$ (c,d). (e) Line-outs of the MCD images along the dashed lines of corresponding colors in (a-d). The dichroic signal is given as a fraction of the peak intensity.}
 \label{fig:SOI MCD}
\end{figure}

For a $\ell_{\text{in}} = 1$ beam incident on a skyrmion, the reflected beam may contain a $\ell=0$ contribution, depending on the incident SAM. Inspecting Fig.\,\ref{fig:simu SOI}.a, we observe that the absence of optical singularity in the $p$ component leads to a Gaussian-like on-axis intensity profile, whereas the presence of an optical singularity results in a very weak on-axis intensity (Fig.\,\ref{fig:simu SOI}.b). Thus, a MCD measurement, consisting in subtracting the far field images obtained with $s_{\text{in}}=1$ and $s_{\text{in}}=-1$, performed with a $\left|\ell_{\text{in}}\right|=1$ incident OAM, should exhibit a measurable on-axis signal. Fig.\,\ref{fig:SOI MCD} shows the predicted dichroic images, for $\ell_{\text{in}}=\pm 1$ incident OAM.

All images are null on-average. Yet, as predicted, they exhibit an on-axis positive signal for $\ell_{\text{in}}=1$ and negative signal for $\ell_{\text{in}}=-1$ incident polarization (see the line-outs in Fig.\,\ref{fig:SOI MCD}.c), independently of the magnetic skyrmion helicity, as the OSOI is independent of the sign of $m_0$ (Eq.\,\ref{eq:mean OAM variation vortex CP}). The peak amplitude of the signal is strong: at maximum about 0.93 times the peak intensity of the reflected beam, and between 30\% and 50\% on the optical axis. The fact that the strongest signal is not observed on axis is due to the reflected beam being a non-pure sum of LG modes, as mentioned in the previous section. MCD images thus provide an experimental observable of OSOI. Contrary to the magnetic helicoidal dichroism images reported in Ref.\,\cite{Fanciulli2022}, here the OSOI signature is not contained in the overall shape of the image, but in its net value - positive or negative - in a localized region of space. %However, the transverse modulation of the MCD images also carries information. It is clear in Fig.\,\ref{fig:SOI MCD} that the left/right antisymmetry of the image is exchanged when switching the helicity $h$ of the Bloch skyrmion. For a Néel skyrmion, one rather observes an up/down antisymmetry (supp mat). The MCD or MHD measurements could thus be used as a probe of the skyrmion type. 

\section*{Discussion}

We identified a form of OSOI occurring in magneto-optics with structured magnetic samples. Our study focused on reflection off a magnetic interface, described by the MOKE formalism. We have shown that the SAM of the incident field affects the OAM content and the average OAM quantum number of the reflected field. We interpret this effect as an interference between the transverse, longitudinal and polar MOKE effects, coupled by the electric field's components of the incident light. This modification of the OAM of the light is not accompanied by an opposite change of the SAM, that would yield a null total angular momentum modification. Therefore, part of light's angular momentum is coherently transferred to matter. How this transfer exactly takes place in our case -- %is it a microscopic mechanism modifying the angular momentum of the electrons in the solid, or 
is a mechanical torque exerted on the texture? -- will have to be investigated in future works. Conversely, the Faraday effect \cite{Schatz1969}, wherein the incident beam is transmitted by a magnetic sample, should allow to observe full SAM to OAM transfer, since, in that case, the sample could be invariant under rotation about the beam axis.

We argue that an experimental setup similar to the one reported in Refs.\,\cite{Fanciulli2022, Fanciulli2025} could be used to observe this OSOI effect by performing MCD measurements. 
Our work allows to anticipate that magnetic textures with appropriate topology could serve as re-configurable OAM beam shapers, similarly to q-plates and metasurfaces in the visible domain. Compared to these passive devices, magnetic textures present the promising advantage of being controllable by external magnetic or electric fields, and exhibit a fast dynamical response. In addition, the conclusions of our study lets us envision the implementation of skyrmion readers based on the SAM/OAM of light. Conversely, one could imagine shaping or controlling magnetic skyrmions with beams of light carrying SAM and/or OAM \cite{Yang2018, Polyakov2020, Guan2023}, with applications in data encoding and processing \cite{Gobel2021}. Finally, the capability of free-electron lasers \cite{Huang2021} or high harmonic generation sources \cite{Geneaux2016, Gauthier2016, Kong2017, Luttmann2023} to synthesise ultrashort femtosecond or attosecond light pulses carrying various kinds of angular momentum lets us envision to perform the MCD experiment described here at the natural time-scales of spin dynamics in matter.

\bibliography{sample}

\section*{Funding}

This work was supported by the Agence Nationale de la Recherche (France), project HELIMAG, Grant No. ANR-21-CE30-0037, and project TORNADO, Grant No. ANR-23-EXLU-0004;Indo-French CEFIPRA Grant Project No.- (7104-I), and by the European Union "Pathfinder" program, Obelix project, Grant agreement ID: 101129641. The authors acknowledge support from the Programm ERC CZ, Project TWISTnSHINE, grant No. LL2314; and from the COST Action NEXT (European Cooperation in Science and Technology, grant n. CA22148).

\section*{Author contributions}

TR, ML and MF first identified the OSOI. ML, TR, MF, PC and MS conducted the analytical calculations. ML and MF performed the numerical simulations. All authors reviewed the manuscript. 

\section*{Competing interests}

The authors have no conflict of interest to disclose.

%The corresponding author is responsible for submitting a \href{http://www.nature.com/srep/policies/index.html#competing}{competing interests statement} on behalf of all authors of the paper. This statement must be included in the submitted article file.


\end{document}